\documentclass[conference]{IEEEtran}

\usepackage{cite}
\usepackage{amsmath,amssymb,amsfonts}

\usepackage[T1]{fontenc}

\usepackage{graphicx}
\usepackage{textcomp}
\usepackage{xcolor}
\def\BibTeX{{\rm B\kern-.05em{\sc i\kern-.025em b}\kern-.08em
    T\kern-.1667em\lower.7ex\hbox{E}\kern-.125emX}}

\usepackage{makecell}
\usepackage{booktabs}
\newcolumntype{Y}{>{\centering\arraybackslash}X}
\usepackage{subcaption}
\usepackage{tikz} 
\usetikzlibrary{calc, arrows.meta, intersections, backgrounds, fit, positioning, shapes.misc, fadings, through,decorations.pathreplacing, shapes,arrows,shapes.multipart, shapes.geometric}
\usetikzlibrary {patterns,patterns.meta} 
\usepgflibrary{patterns.meta} 
\usepackage{multirow}
\usepackage{tabularx}
\usepackage{pgfplots, pgfplotstable} 
\usepackage{bm}
\usepackage{array}
\usepackage{url}

\usepackage{algorithm}
\usepackage{algpseudocode}

\tikzset{
decision/.style = {diamond, draw, fill=blue!20, font = \scriptsize, text width=2.5em, text centered, node distance=3cm, inner sep=0pt},
decision2/.style = {diamond, draw, aspect = 2, fill=blue!20, font = \scriptsize, text width=2.5em, text centered, node distance=3cm, inner xsep=1pt}, 
block/.style = {rectangle, draw, fill=blue!20, 
  text width=4em, 
  font = \scriptsize,
  text centered, 
  rounded corners, minimum height=3em, execute at begin node={\hyphenpenalty=10000\exhyphenpenalty=10000\relax}}, 
blocknor/.style = {rectangle, draw, fill=blue!20,  text width=4em, 
  font = \scriptsize, text centered,  minimum height=3em, execute at begin node={\hyphenpenalty=10000\exhyphenpenalty=10000\relax}}, 
octagon/.style = {shape=regular polygon, regular polygon sides=8, font = \scriptsize, draw, inner sep = 0pt, text centered, execute at begin node={\hyphenpenalty=10000\exhyphenpenalty=10000\relax}},
line/.style = {draw, -latex'},
biline/.style = {draw, latex'-latex'},
cloud/.style = {draw, ellipse,fill=red!20, node distance=3cm, minimum height=2em},
subroutine/.style = {draw,rectangle split, rectangle split horizontal, rectangle split parts=3,minimum height=1cm, rectangle split part fill={red!50, green!50, blue!20, yellow!50}},
mycircle/.style = {draw, circle, fill=blue!20, text centered, font = \scriptsize, execute at begin node={\hyphenpenalty=10000\exhyphenpenalty=10000\relax}},
connector/.style = {draw,circle,node distance=3cm,fill=yellow!20},
data/.style = {draw, trapezium,node distance=3cm,fill=olive!20},
picturebox/.style={draw,rounded corners,minimum width=3.32cm,minimum height=2 cm,align=center},  
FIT/.style = {draw, fill=gray!30, inner xsep=4mm, inner ysep=3mm, fit=#1},        
closeFIT/.style = {draw, 
                 inner xsep=1mm, inner ysep=1mm, fit=#1},
dashFIT/.style = {draw, dashed, 
                 inner xsep= 2mm, inner ysep=1mm, fit=#1},
closeFITnoline/.style = {inner xsep=1mm, inner ysep=1mm, fit=#1},
person/.style={trapezium, draw, trapezium stretches=true, trapezium left angle=75, trapezium right angle=75, 
    align=center, inner sep = 2pt, font = \scriptsize, text centered, execute at begin node={\hyphenpenalty=10000\exhyphenpenalty=10000\relax}},
person2/.style={ chamfered rectangle, draw, chamfered rectangle xsep=3mm, chamfered rectangle ysep=1mm, align=center, inner sep=1pt, font=\scriptsize, text centered, execute at begin node={\hyphenpenalty=10000\exhyphenpenalty=10000\relax}}
}

\newcommand{\projname}{LaDA}
\newcommand{\mytextsc}[1]{{\fontfamily{ptm}\selectfont\textsc{#1}}}

\title{\mytextsc{\projname}: Reverse Engineering Registers through Layout and Dataflow Analysis
}

\author{Kevin Xu and Samuel Pagliarini \\
Carnegie Mellon University \\ \{kx2, pagliarini\}@cmu.edu}

\pgfplotsset{compat=1.18} 

\begin{document}

\maketitle

\begin{abstract}

Register reverse engineering (RE) aims to reconstruct register groupings from individual flip-flops (FFs) in a gate-level netlist, providing valuable structural information for hardware security. Existing register RE techniques mainly utilize netlist features to build their heuristics while disregarding physical cell locality data. We propose \mytextsc{\projname}, a register RE technique that integrates dataflow analysis with cell locality information to improve the quality of the register grouping. \mytextsc{\projname} introduces a processing filter that rejects merge and split operations, based on cell proximity, followed by a postprocessing method that combines netlist properties, including predecessor-successor relationships and control signals, with cell coordinate-based features. Our work is the first technique to consider cell coordinates for ASIC register RE. \mytextsc{\projname} achieves an average normalized mutual information score (NMI) of 0.95, a 10.8\% improvement over best known results from \mytextsc{DANA}, a well-established register RE method in the literature. 

 \end{abstract}


\section{Introduction}


The globalized IC supply chain is composed of various third-party fabs, design houses, and other entities that form a highly interdependent ecosystem. However, this interdependence introduces security risks such as hardware Trojan insertion, IP theft, and overproduction~\cite{azriel2021survey}. Counterfeiting and overproduction alone are estimated to cost U.S. semiconductor companies more than \$7.5 billion annually~\cite{SIA2018counterfeit}.

Logical reverse engineering (RE) has emerged as a potential approach for addressing these security risks, by algorithmically identifying high-level components, such as registers or finite state machines, from low-level representations, such as gate-level netlists~\cite{xu2026survey}. The recovery of these components supports applications in hardware Trojan insertion/detection~\cite{re_for_troj_insertion} and IP piracy detection, and can also aid in the reconstruction of a chip's design structure and functionality. Logical RE should be distinguished from physical RE, which involves decapsulating, delayering, and imaging a fabricated chip to recover its gate-level netlist~\cite{torrance2009state}. Physical RE is a prerequisite for logical RE, but it is a distinct, well-established procedure with a mature workflow. Logical RE, on the other hand, is a nascent field that faces a number of complex research challenges~\cite{xu2026survey}. 

A primary challenge in logical RE is overcoming the opaque, flattened nature of gate-level netlists, which are deprived of the original design's structure and hierarchy. To identify high-level modules in this ``sea of gates,'' logical RE techniques commonly transform the netlist into a graph, where the gates become nodes and the wires become edges; this representation permits the application of established graph analyses, such as segmentation and clustering, along with circuit-specific strategies, including fan-in cone analysis and subgraph matching~\cite{he2021arithmetic, li2025dagnnre}. 

However, existing logical RE techniques generally treat the gate-level netlist as the sole source of circuit information. In a realistic RE scenario, the artifact examined through physical RE is the chip itself. From this examination, often performed layer by layer, spatial and geometric information can be obtained, information that is otherwise absent in netlists. This includes metal layers, pins, vias, as well as the locations (coordinates) and orientations of individual logic cells.

We hypothesize that locality data can augment logical RE techniques through the introduction of cell proximity (or distance) as a feature. Specifically, \mytextsc{\projname} applies layout locality features to register RE, the goal of which is to cluster flip-flops (FFs) into their original register groupings; accurate register identification can improve the understanding of an IC's dataflow and reveal the register-transfer relationships at the heart of the design. We make the following contributions:



\begin{itemize}
    \item We create a database of gate-level netlists and corresponding auxiliary files containing cell coordinates.
    \item We develop the first ASIC-specific layout-aware logical RE method by combining FF spatial proximity with netlist structural features to improve register grouping. 
    \item We open source our findings at~\cite{anonrepo}, including the source code for \mytextsc{LaDA}, netlists, and corresponding layouts. 
\end{itemize}


\section{Background}

Register RE encompasses a wide range of methodologies, often relying on structural and functional netlist features to group FFs into their original registers. One class of techniques~\cite{shi2010efficient, subramanyan2013funct, fyrbiak2019difficulty, rao2024register} groups sequential elements driven by the same control signals, which are defined by the nets directly driving the enable, (re)set, or clock pin of a FF. However, synthesis optimizations can disrupt the fan-in signal patterns of a register. Hence, this heuristic is often applied as a preprocessing step but not as a standalone method. For instance,~\cite{subramanyan2014structfunct} first performed control signal matching, then searched for specific arrangements of FFs, multiplexers, and decoders to identify CPU register files and data registers; these structural definitions are too narrow for general use. 


Another line of research focuses on the identification of datapath words, which intersects with register RE because word wires often correspond to register fan-outs. This approach also introduced fan-in analysis, which examines the logic leading to a target instance. An influential early technique, \mytextsc{WordRev}~\cite{li2013wordrev}, grouped wires with structurally or functionally equivalent fan-ins into words. Subsequent work removed control signals from fan-in cones to improve identification accuracy~\cite{tashjian2015control}. In a different approach,~\cite{yu2016word} converted netlists into algebraic expressions for functional analysis.

Fan-in analysis regained prominence with \mytextsc{RELIC}~\cite{meade2016relic}, an important work that classified registers as datapath or control logic using fan-in similarity scores. Several follow-ups reduced \mytextsc{RELIC}'s runtime~\cite{brunner2019fastrelic}, and analyzed the Boolean functions of fan-in slices~\cite{geist2020relicfun}. Subsequent techniques in the ``\mytextsc{RELIC} family,'' namely \mytextsc{REBUS}, \mytextsc{REWIND}, and \mytextsc{REPCA}, shifted the focus toward word identification~\cite{meade2016toolsetrebus, meade2018old}. The technique in~\cite{meade2018old} also introduced normalized mutual information (NMI) to measure grouping quality, advancing standardized evaluation in RE studies. As machine learning gained popularity, \mytextsc{ReIGNN} and \mytextsc{RELIC-GNN} applied graph neural networks to register classification~\cite{chowdhury2021reIGNN, dong2023relicGNN}, and \mytextsc{ReBERT}~\cite{zhang2025llm} used large language models to group bits into words. 


The main limitation of the ``\mytextsc{RELIC} family'' methods (particularly~\cite{meade2016relic, brunner2019fastrelic, geist2020relicfun}) is their dependence on manually-set parameters, especially the fan-in traversal depth used to compute similarity scores. Excessive depth could capture irrelevant circuitry and increase runtime, while insufficient depth could omit useful structural information. Selecting an appropriate depth therefore requires time-consuming tuning and may vary across netlists.

To address this limitation, \mytextsc{DANA}~\cite{albartus2020dana} analyzed the FF dataflow. Its preprocessing stage filtered out combinational logic, producing an FF dependency graph. The processing stage iteratively grouped FFs sharing the same clock signals or common predecessor and successor FF sets (which represent the fan-in and fan-out relationships, respectively, of a register). By relying on FF dataflow, this method avoided the previous issues with ``magic numbers,'' netlist-dependent parameter fine-tuning, and common control signal matching. 

Nevertheless, \mytextsc{DANA} utilizes only netlist structural features, and prior ASIC logical RE approaches did not consider layout features to improve RE accuracy. A recent technique,~\cite{venkatesan2023layout}, used layout location data to reverse engineer combinational sub-circuits, but it targeted FPGAs rather than ASICs. The method combined Manhattan distance and graph edit distance, with a weight function of site, tile, and clock region features; the site and tile features are specific to FPGAs. It also used a fixed spatial distance threshold of 100 units, limiting generalizability across designs with different sizes/scales; this limitation is particularly concerning for ASICs, which have chip-specific floorplan dimensions. Moreover, most benchmarks in~\cite{venkatesan2023layout} contained fewer than 10,000 cells, leaving the scalability of layout-based RE relatively unexplored. 
 
\begin{figure}
    \centering 
    \includegraphics[width=0.5\textwidth, trim={0cm 0.3cm 0cm 0.5cm}, clip]{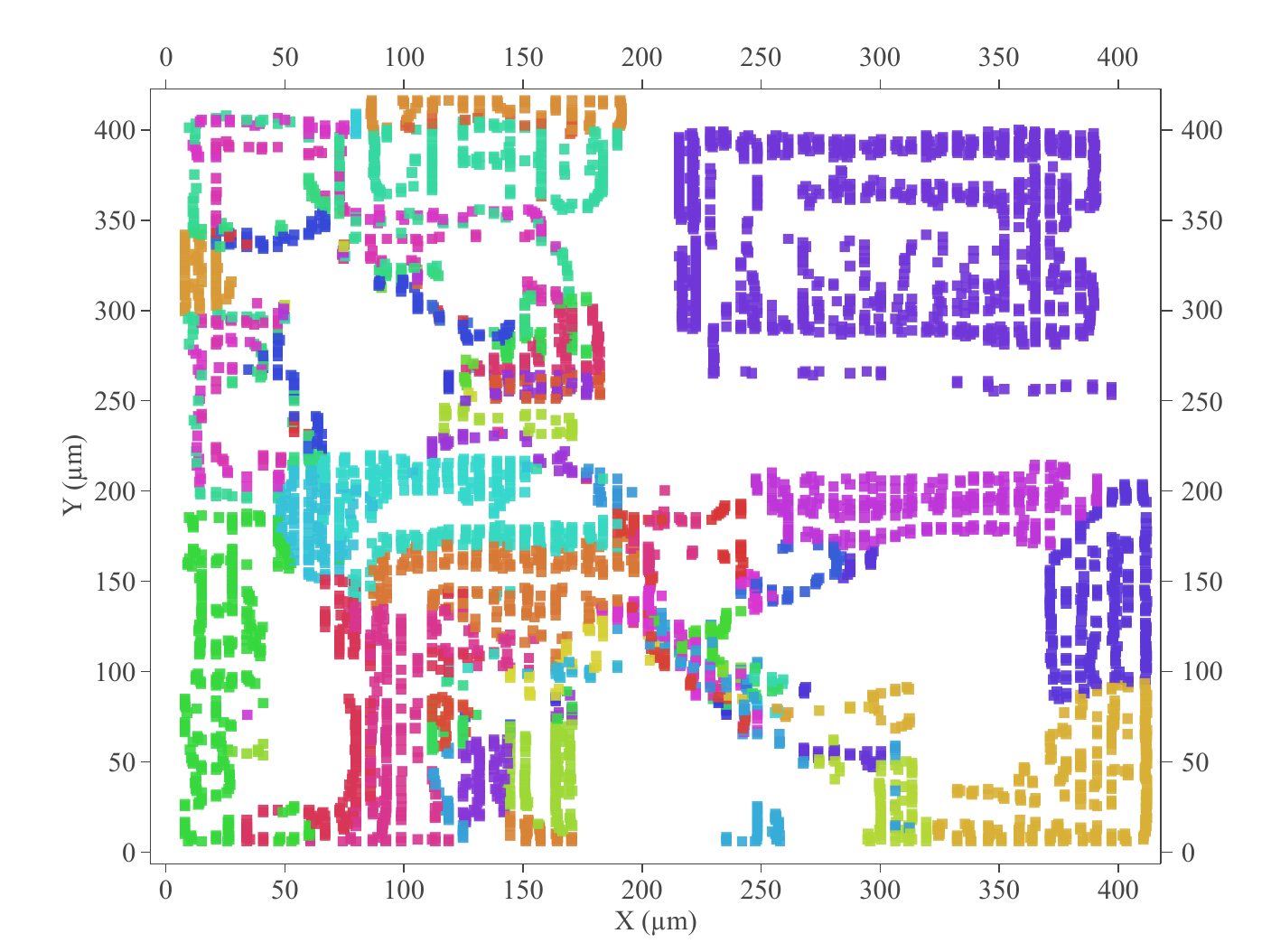}
    \caption{Plot of layout FF cell coordinates for a RISC-V CPU. Different register groups are highlighted in different colors.}
    \label{fig:layout}
\end{figure}

\textbf{Novelty of our solution}: \mytextsc{\projname} integrates a layout cell coordinate-based algorithm with \mytextsc{DANA}'s netlist structural features, and harnesses indirect, nearest-neighbor methods to measure layout proximity; unlike methods based on fixed physical distance thresholds, \mytextsc{\projname} generalizes across register groups of varying sizes and spatial distributions. The main insight is that cells belonging to the same module or register are often placed close together to reduce interconnect length, and improve timing and power efficiency. Fig.~\ref{fig:layout} plots the FF cell coordinates over the layout of a RISC-V CPU, demonstrating the proximity of FFs within the same register and the irregular register shapes and distributions that make register RE challenging.

\section{Layout-enhanced Reverse Engineering}


The proposed \mytextsc{\projname} method consists of two stages: the first augments \mytextsc{DANA}'s processing stage by rejecting groupings of spatially distant registers, and the second is a novel postprocessing algorithm that merges and splits \mytextsc{DANA}'s register groups, using FF cell coordinate distance as a major heuristic, alongside netlist features such as successor-predecessor (S-P) set similarity. A visualization of the method is shown in Fig.~\ref{fig:pipeline}.

\subsection{Processing Filter}

\mytextsc{DANA}'s processing phase repeatedly executes combinations of nine ``passes'' until a final grouping is decided by majority voting. Each pass takes an input grouping of registers, splits or merges different groups based on a shared netlist property, and outputs a new register grouping. 

\mytextsc{\projname} extends this processing phase by filtering the merge and split passes based on spatial proximity. Specifically, the proposed filter blocks a merge pass if the candidate groups are spatially distant, and conversely, blocks a split if the groups are spatially proximate. Group locations are approximated using bounding boxes, whose corner coordinates are updated as FFs are added to or removed from the candidate groups. The distance between two groups is estimated by the edge-to-edge distance between their bounding boxes, with a threshold $T_1$ defined as a ratio of the floorplan width. 


We first evaluated the filter on the nine benchmark netlists considered in~\cite{albartus2020dana}. The filter yielded substantial NMI improvements for two designs (SHA3 and AES). We also tested independent merge and split passes relying solely on cell distances. However, these na\"ive strategies sometimes performed unnecessary merges or splits, leading to clustering mistakes. Overall, the initial results are encouraging, but these experiments motivate a postprocessing strategy that provides finer control over grouping changes.


\newcommand{\blockh}{1.5em} 
\newcommand{\rectw}{6em}
\newcommand{\sblockh}{2.2em} 
\newcommand{\sblockw}{3.5em}
\newcommand{\shblockw}{3em}

\newcommand{\danafill}{purple!15}
\newcommand{\danabord}{purple!70!black}
\newcommand{\mergefill}{cyan!15}
\newcommand{\mergebord}{cyan!70!black}
\newcommand{\splitfill}{orange!15}
\newcommand{\splitbord}{orange!80!black}

\newcommand{\verticaldist}{0.3cm}
\newcommand{\shortvert}{0.12cm}
\newcommand{\horizdist}{0.2cm}
\newcommand{\shorthoriz}{0.1cm}
\newcommand{\legendlen}{0.3cm}
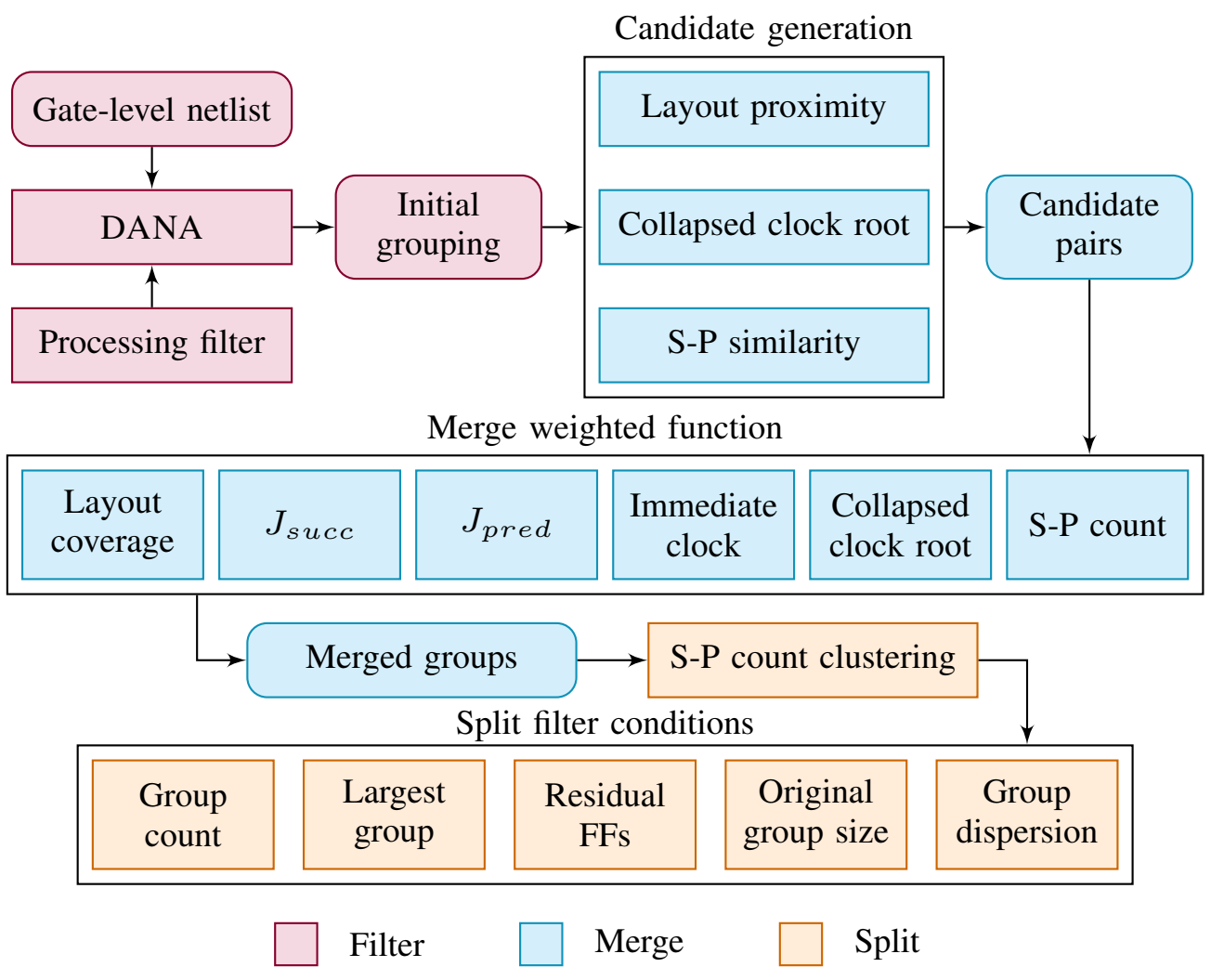
\begin{figure}
    \centering
    \begin{tikzpicture} [node distance = 2.5 cm, auto]      
        \node[block, draw = \danabord, fill = \danafill, minimum height = \blockh, text width = \sblockw] (grouping) {Initial grouping};
        \node[blocknor, draw = \danabord, fill = \danafill, minimum height = \blockh, text width = \rectw-1em, left = 1.5*\horizdist of grouping] (dana) {\mytextsc{DANA}};
        \node[block, draw = \danabord, fill = \danafill, minimum height = \blockh, text width = \rectw-1em, above = \verticaldist of dana] (netlist) {Gate-level netlist};
        \node[blocknor, draw = \danabord, fill = \danafill, minimum height = \blockh, text width = \rectw-1em, below = \verticaldist of dana] (filter) {Processing filter};
        
        \node[blocknor, draw = \mergebord, fill = \mergefill, minimum height = \blockh, text width = \rectw, right = 2*\horizdist of grouping] (can2) {Collapsed clock root};
        \node[blocknor, draw = \mergebord, fill = \mergefill, minimum height = \blockh, text width = \rectw, above = \verticaldist of can2] (can1) {Layout proximity};
        \node[blocknor, draw = \mergebord, fill = \mergefill, minimum height = \blockh, text width = \rectw, below = \verticaldist of can2] (can3) {S-P similarity}; 
        \node[block, draw = \mergebord, fill = \mergefill, minimum height = \blockh, text width = \sblockw, right = 2*\horizdist of can2] (candidates) {Candidate pairs};

        \node[below = 2*\verticaldist of filter, minimum height = \sblockh, text width = 3em] (belowcand) {};
        \node[blocknor, draw = \mergebord, fill = \mergefill, minimum height = \sblockh, text width = \shblockw, below right = -1*\sblockh and -4.5em of belowcand] (feat1) {Layout coverage};
        \node[blocknor, draw = \mergebord, fill = \mergefill, minimum height = \sblockh, text width = \shblockw, right = \shorthoriz of feat1] (feat2) {$J_{succ}$};
        \node[blocknor, draw = \mergebord, fill = \mergefill, minimum height = \sblockh, text width = \shblockw, right = \shorthoriz of feat2] (feat3) {$J_{pred}$};
        \node[blocknor, draw = \mergebord, fill = \mergefill, minimum height = \sblockh, text width = \shblockw, right = \shorthoriz of feat3] (feat4) {Immediate clock};
        \node[blocknor, draw = \mergebord, fill = \mergefill, minimum height = \sblockh, text width = \shblockw, right = \shorthoriz of feat4] (feat5) {Collapsed clock root};
        \node[blocknor, draw = \mergebord, fill = \mergefill, minimum height = \sblockh, text width = \shblockw, right = \shorthoriz of feat5] (feat6) {S-P count};        

        \scoped[on background layer]
        \node[fill = white] (candback) [closeFIT = (can1) (can2) (can3)] {}; 
        \node[above= -0.06cm of candback, font = \scriptsize] (candlabel) {Candidate generation};

        \scoped[on background layer]
        \node[fill = white] (mergeback) [closeFIT = (feat1) (feat2) (feat3) (feat4) (feat5) (feat6)] {}; 
        \node[above= -0.06cm of mergeback, font = \scriptsize] (mergelabel) {Merge weighted function};

        \node[block, draw = \mergebord, fill = \mergefill, minimum height = \blockh, text width = \rectw, below right = \verticaldist and 0.3cm of feat1] (mergegroups) {Merged groups};

        \node[blocknor, draw = \splitbord, fill = \splitfill, minimum height = \blockh, text width = \rectw, right = 2.5*\horizdist of mergegroups] (labelsp) {S-P count clustering};

        \node[blocknor, draw = \splitbord, fill = \splitfill, minimum height = \sblockh, text width = \shblockw, below = 3.9*\verticaldist of mergeback.south] (filter3) {Residual FFs};
        \node[blocknor, draw = \splitbord, fill = \splitfill, minimum height = \sblockh, text width = \shblockw, left = \horizdist of filter3] (filter2) {Largest group};
        \node[blocknor, draw = \splitbord, fill = \splitfill, minimum height = \sblockh, text width = \shblockw, left = \horizdist of filter2] (filter1) {Group count};
        \node[blocknor, draw = \splitbord, fill = \splitfill, minimum height = \sblockh, text width = \shblockw, right = \horizdist of filter3] (filter4) {Original group size};
        \node[blocknor, draw = \splitbord, fill = \splitfill, minimum height = \sblockh, text width = \shblockw, right = \horizdist of filter4] (filter5) {Group dispersion};        

        \scoped[on background layer]
        \node[fill = white] (splitback) [closeFIT = (filter1) (filter2) (filter3) (filter4) (filter5)] {}; 
        \node[above= -0.1cm of splitback, font = \scriptsize] (splitlabel) {Split filter conditions};

        \path[line] (netlist) -- (dana);
        \path[line] (filter) -- (dana);
        \path[line] (dana) -- (grouping);
        \path[line] (grouping.east) -- (candback.west);
        
        \path[line] (candback.east) -- (candidates);        
        \path[line] (candidates.south) -- (candidates.south |- mergeback.north);

        \coordinate (mergestart) at ([xshift=0.6cm]feat1.south);
        \path[line] (mergestart |- mergeback.south) |- (mergegroups.west);
        \path[line] (labelsp.east) -| (filter5.north |- splitback.north);
        \path[line] (mergegroups) -- (labelsp);
    \end{tikzpicture}

    \begin{tikzpicture}[baseline=(current bounding box.center), node distance=0.15cm]
        \node[draw=\danabord, fill=\danafill, minimum width=\legendlen, minimum height=\legendlen, inner sep=0pt] (leg1) {};
        \node[right=0.10cm of leg1, font=\scriptsize] (txt1) {Filter};
    
        \node[draw=\mergebord, fill=\mergefill, minimum width=\legendlen, minimum height=\legendlen, inner sep=0pt, right=0.55cm of txt1] (leg2) {};
        \node[right=0.10cm of leg2, font=\scriptsize] (txt2) {Merge};
    
        \node[draw=\splitbord, fill=\splitfill, minimum width=\legendlen, minimum height=\legendlen, inner sep=0pt, right=0.55cm of txt2] (leg3) {};
        \node[right=0.10cm of leg3, font=\scriptsize] (txt3) {Split};
    \end{tikzpicture}
    \caption{\mytextsc{\projname} algorithm flowchart}
    \label{fig:pipeline}
    \vspace{-1em}
\end{figure}

\subsection{Postprocessing Algorithm}

The initial evaluation revealed that spatial proximity alone is not always an effective predictor of register groupings. We initially expected FFs from the same register to be placed close together and different registers to form distinct clusters. Although many registers follow this pattern, certain ``edge cases'' may overlap the same area as other registers, especially when they share a similar functionality and must be positioned near the same combinational logic cloud. For example, the purple and green registers at the top left of Fig.~\ref{fig:layout} overlap.

In other cases, FFs within one register can be dispersed over a wide swathe of the layout, while those of other registers are tightly packed, making a single distance threshold inadequate. For example, in Fig.~\ref{fig:layout}, the purple and green registers at the bottom left have a noticeable size disparity. Conventional clustering algorithms can also struggle because different registers are often adjacent or overlapping; if cell coordinates are the only features, the FFs will likely be clustered into a single, all-encompassing register. In Fig.~\ref{fig:layout}, only the purple register at the top right and the blue register at the bottom are distinctly isolated; the other registers are all adjacent or overlapping.

Therefore, our proposed postprocessing method integrates layout proximity with established netlist structural features. This method consists of two phases: the first phase merges registers that were fractured into multiple groups, and the second phase splits registers incorrectly merged by \mytextsc{DANA} (or the first phase).

\subsubsection{Merge}
\label{sec:merge}
In the merge phase, the method generates pairs of candidate groups, scores each pair using a weighted function of layout and netlist features, and merges the strongly connected groups (high-scoring pairs). 

Candidate generation consists of three approaches. The first approach forms pairs from spatially proximate groups. It constructs an R-tree of the FF cell coordinates, using the Boost library implementation~\cite{boost}; an R-tree organizes objects in minimum bounding rectangles and is well-suited for spatial data. Next, the approach iterates over all FFs, pairing each FF's group with the groups containing its $k$-nearest neighbor FFs in the R-tree. Importantly, this method \textbf{removes} the need for a design-dependent distance threshold. 

The second approach addresses clock buffering (or useful skew), which can cause FFs in the same clock domain and register to have different clock nets. As a solution, the approach searches backwards from each group's clock signal, until reaching an instance that is neither an inverter nor a buffer; the output net at this point is defined as the \emph{collapsed clock root}. Then, each group is paired with the $g$ groups that share the same clock root and have the nearest centroids by layout distance.

Finally, the third approach assumes that FFs in the same register have similar sets of predecessor and successor registers. Rather than requiring exact S-P matches, as in \mytextsc{DANA}, it measures set similarity, using the Jaccard index~\cite{jaccard1901index}. For each pair of groups, predecessor and successor similarities are computed as $J_{pred}$ and $J_{succ}$, respectively, with overall similarity defined as $min(J_{pred}, J_{succ})$. Candidate pairs are formed when each group ranks the other among its $l$-most similar groups. The groups must also share the same immediate clock net, a requirement to keep the clustering conservative.


Candidate pair scores are computed as a weighted sum of six layout and netlist features. The first and second features are predecessor and successor similarity, again measured by the Jaccard index. The third feature is layout coverage, defined as the fraction of FFs in each group that appear among the physical $k$-nearest neighbors of FFs in the other group; its mathematical expression is shown in (\ref{eq:coverage}), where $G_i$ and $G_j$ are the two candidate groups, and $N_i^{(j)}$ denotes the set of FFs in group $G_i$ that have a $k$-nearest neighbor relationship with an FF in group $G_j$. 

\begin{equation}
L_{ij} =
\min\left(
\frac{|N_i^{(j)}|}{|G_i|},
\frac{|N_j^{(i)}|}{|G_j|}
\right)
\label{eq:coverage}
\end{equation}

The fourth feature is immediate clock equality, a binary indicator of whether the two groups have the same clock net. The fifth feature is collapsed clock root similarity, defined as the Jaccard index between the groups' sets of collapsed clock roots. The sixth feature is connectivity-count similarity, a comparison of the average numbers of predecessors and successors for each candidate group. The feature calculations are given in

\begin{equation}
\begin{gathered}
\mathbf{q}_i =
\Bigl(
\frac{1}{|G_i|}\sum_{g\in G_i}\ln(1+p_g),
\frac{1}{|G_i|}\sum_{g\in G_i}\ln(1+s_g)
\Bigr)
\\
C_{ij}^{\mathrm{count}} =
\exp\left(-\left\|\mathbf{q}_i-\mathbf{q}_j\right\|_2\right).
\end{gathered}
\label{eq:count_similarity}
\end{equation}

\noindent where $G_i$ denotes a candidate register group, \emph{g} is an FF in $G_i$, $p_g$ and $s_g$ are the numbers of predecessor and successor FFs for \emph{g}, respectively, and $q_i$ is the resulting S-P count feature of $G_i$. The connectivity-count similarity $C_{ij}^{\mathrm{count}}$ is then computed by applying exponential decay to the Euclidean distance between the \emph{q} values of the two candidate groups. 

All group pairs with scores above threshold $T_2$ are selected as merge candidates. In the final step, the merge method clusters all groups connected by high-scoring pairs into registers through a union-find data structure. 

\subsubsection{Split}
The split phase searches the input grouping for distinct populations of FFs within each candidate group. Since aggressive splitting can dramatically reduce grouping quality, \mytextsc{\projname} employs conservative criteria. The split method uses similar S-P counts, rather than set similarity. Each FF \emph{g} is assigned a structural identifier $\sigma (g) = (p_g, s_g)$, where $p_g$ and $s_g$ are its numbers of predecessors and successors. FFs with identical identifiers form an initial grouping, after which groups whose $\sigma (g)$ distance is below a threshold $T_3$ are connected. A union-find data structure then merges the connected components into the output register groups.  

Additional criteria reject a split if the final set contains more than $C_1$ groups (filtering excessively high fragmentation), if the largest group contains at least $C_2$ (a proportion in $[0,1]$) of the original group's FFs, or if residual, ungrouped FFs account for at least $C_3$ of the original group's FF count (i.e., too many FFs remain ungrouped). In addition, the split method ignores groups containing fewer than $C_4$ FFs, assuming that smaller groups are less likely to result from incorrect mergers.

The last criterion rejects a split if the child groups are widely dispersed. A preprocessing step computes the median distance from each FF to its nearest FF, denoted as $D_{FF}$. After a given group is split into child groups, the method calculates their centroids and $D_{child}$, the median pairwise distance between them. The split is accepted only if $D_{child} / D_{FF} \leq C_5$. This criterion handles the edge case in which FFs from different registers are overlapping or adjacent, while avoiding fragmentation of large, spatially dispersed registers. The condition is not applied to groups without successors, which are treated as likely output registers. Such registers may have FFs dispersed along the floorplan edges and are therefore more susceptible to false merging.

\subsection{Benchmarking and Evaluation}


To compare the performance of \mytextsc{\projname} and \mytextsc{DANA}, we evaluate their register groupings against the ground truth using NMI and purity. Purity measures the homogeneity of predicted groups with respect to the ground truth, while NMI measures the overall agreement between predicted groups and ground truth groups; both range from 0 to 1~\cite{strehl2003cluster}. As in \mytextsc{DANA}, the ground truth is derived from RTL register instance names. 





\mytextsc{\projname} and \mytextsc{DANA} are evaluated on two benchmark datasets. The first dataset is \mytextsc{DANA}'s own benchmark suite, which comprises nine designs, spanning cryptographic cores, processors, and system-on-chips. The second dataset consists of 60 open-source designs drawn from OpenCores~\cite{opencores} and various repositories, covering a wide range of accelerators. The initial thresholds and parameters of \mytextsc{\projname} were determined through tests on the \mytextsc{DANA} dataset. The second dataset is included to assess potential overfitting.

We synthesize the designs' RTL using Synopsys DC v23.12, generate the corresponding layouts using Cadence Innovus v21.16, and extract FF cell coordinates using a heavily modified version of the ``lefdef'' DEF parser~\cite{lefdef}. Since \mytextsc{DANA}'s benchmarks were synthesized with clock gating, and its processing algorithm considers clock nets as control signal features, we hypothesize that the performance of \mytextsc{DANA} degrades in the absence of clock gating. Hence, every design is synthesized twice, with and without clock gating.

Another important synthesis consideration is the standard cell library. \mytextsc{DANA} uses LSI10k, an academic library that lacks the physical library and technology files necessary for layout generation. We instead use the ASAP7 open-source PDK~\cite{asap7} that contains both a standard cell library and its physical design files. Without loss of generality, we consider results collected from the ASAP7 netlists as the baseline for comparison with \mytextsc{\projname}, rather than the values reported in~\cite{albartus2020dana}. \mytextsc{\projname} is developed in C++ and extends the \mytextsc{DANA} codebase as a plugin for HAL~\cite{wallat2019HAL}. 

\mytextsc{\projname}'s parameters are summarized in Table~\ref{tab:parameters}. The thresholds, split criteria, candidate generation values, and merge function feature weights follow the order in which the corresponding methods and features are introduced. The thresholds were determined empirically through testing on the \mytextsc{DANA} benchmark set. All parameter values are held constant throughout the experiments. Our results show that the parameters generalize well for a set of 60 previously unseen designs.







\begin{table}[ht]
\centering
\caption{\textsc{\projname} Parameters}
\label{tab:parameters}
\small
\setlength{\tabcolsep}{4pt}
\begin{tabular}{l l}
\toprule
\textbf{Component} & \textbf{Specification} \\
\midrule

\multicolumn{2}{l}{\textit{Empirical thresholds}} \\
\quad $T_1$--$T_3$
& 0.25, 0.75, 0.4 \\

\addlinespace
\multicolumn{2}{l}{\textit{Preset design values}} \\
\quad Split constraints $C_1$--$C_5$
& 6, 0.65, 0.15, 16, 15 \\

\quad Candidate generation
& $k=16$, $g=8$, $l=2$ \\

\quad Merge scoring weights
& 0.25, 0.15, 0.20, 0.10, 0.20, 0.10 \\

\bottomrule
\end{tabular}
\end{table}




\section{Results}
\label{sec:results}


Table~\ref{tab:dana_results} compares the NMI and purity results of the \mytextsc{DANA} baseline method with \mytextsc{\projname} on the original \mytextsc{DANA} benchmarks. The higher NMI or purity values are bolded. The full NMI and purity results for the 60-design benchmark set are omitted in the interest of brevity. The bar graphs in Figs.~\ref{fig:nmi_w_cg} and~\ref{fig:nmi_wo_cg} illustrate the NMI difference between \mytextsc{\projname} and \mytextsc{DANA}, for the two benchmark datasets with and without clock gating, respectively. Due to space constraints, only selected benchmarks with noticeable NMI changes are shown. Green bars denote NMI improvements with \mytextsc{\projname}, and red bars denote NMI decreases.


\begin{table}[t]
\centering
\caption{Register Grouping Results on the \mytextsc{DANA} Benchmark Set}
\label{tab:dana_results}
\begin{tabularx}{\linewidth}{l c *{4}{Y}}
\toprule
\multirow{2}{*}{\textbf{Design}}
& \multirow{2}{*}{\textbf{\#FFs}}
& \multicolumn{2}{c}{\textbf{\mytextsc{DANA}}}
& \multicolumn{2}{c}{\textbf{\mytextsc{\projname}}} \\
\cmidrule(lr){3-4} \cmidrule(lr){5-6}
& & \textbf{NMI} & \textbf{Purity}
& \textbf{NMI} & \textbf{Purity} \\
\midrule
aes       & 6720  & 0.718 & 0.257 & \textbf{0.869} & \textbf{0.652} \\
basicrsa  & 8715  & 0.758 & 0.827 & \textbf{0.930} & \textbf{0.882} \\
des       & 1984  & 0.849 & 0.488 & \textbf{0.929} & \textbf{0.964} \\
edge      & 2929  & \textbf{0.984} & 0.940 & \textbf{0.984} & \textbf{0.942} \\
ibex      & 2303  & 0.962 & \textbf{0.939} & \textbf{0.967} & 0.937 \\
open8     & 208   & 0.943 & \textbf{0.923} & \textbf{0.956} & 0.899 \\
opentitan & 22203 & 0.928 & \textbf{0.867} & \textbf{0.934} & 0.830 \\
present   & 151   & 0.917 & \textbf{1.000} & \textbf{1.000} & \textbf{1.000} \\
sha3      & 2245  & 0.665 & \textbf{1.000} & \textbf{0.993} & 0.999 \\
\bottomrule
\end{tabularx}
\end{table}

\definecolor{positivefill}{RGB}{198,239,206}
\definecolor{positiveedge}{RGB}{0,100,0}
\definecolor{negativefill}{RGB}{255,199,206}
\definecolor{negativeedge}{RGB}{156,0,6}
\newcommand{\bargraphheight}{5.0cm} 
\newcommand{\bargraphheightshort}{4.2cm} 

\begin{figure*}[ht]
\centering
\begin{tikzpicture}
\begin{axis}[
    ybar,
    width=0.79\textwidth,
    height=\bargraphheightshort,
    bar width=5pt,
    ymin=-5,
    ymax=35,
    ylabel={NMI Difference (\%)},
    ylabel style={xshift=4pt},
    xtick={0,...,28},
    xticklabels={aes,basicrsa,des,edge,ibex,open8,opentitan,present,sha3,,,dma\_ahb64,keccak,riscv\_ss\_32,warp,xtea,ae18\_core,aes\_128,altor32\_cpu\_l,avr\_core,caes,dma\_ahb32,mesi\_isc,openmsp430,s13207,s38584,signed\_divider,t6507lp,uart2spi},
    x tick label style={rotate=45,anchor=east,font=\scriptsize},
    ymajorgrids=true,
    grid style={dashed},
    ytick={-5,5,10,15,20,25,30,35},
    extra y ticks={0},
    extra y tick labels={0},
    extra y tick style={grid=major,grid style={solid,black,line width=0.8pt},tick style={draw=none}},
    xtick pos=bottom,
    ytick pos=left,
    axis line style={black},
    tick align=outside,
    enlarge x limits=0.02,
]

\addplot[ybar,bar shift=0pt,fill=positivefill,draw=positiveedge,line width=0.7pt] coordinates {
    (0,15.0466)
    (1,17.2403)
    (2,7.9366)
    (3,0.0560)
    (4,0.4672)
    (5,1.2634)
    (6,0.6074)
    (7,8.3427)
    (8,32.8056)

    (11,6.7091)
    (12,29.5850)
    (13,11.3958)
    (14,6.2859)
    (15,10.7296)

    (16,3.4801)
    (17,12.4276)
    (18,13.8688)
    (19,11.3435)
    (21,5.8993)
    (22,10.3849)
    (23,6.5345)
    (24,4.0021)
    (25,15.5459)
    (26,3.9824)
    (27,3.2525)
    (28,3.7054)
};

\addplot[ybar,bar shift=0pt,fill=negativefill,draw=negativeedge,line width=0.7pt] coordinates {
    (20,-3.0611)
};

\draw[black,densely dashed] (axis cs:9.5,-5) -- (axis cs:9.5,35);


\end{axis}
\end{tikzpicture}
\caption{NMI changes from \mytextsc{DANA} to \mytextsc{\projname}. The left portion of the plot shows the 9 original \mytextsc{DANA} designs; the right portion shows only benchmarks with $|\Delta \mathrm{NMI}| > 3\%$. All designs are synthesized with automatic clock gating \textbf{enabled}.}
\label{fig:nmi_w_cg}
\end{figure*}
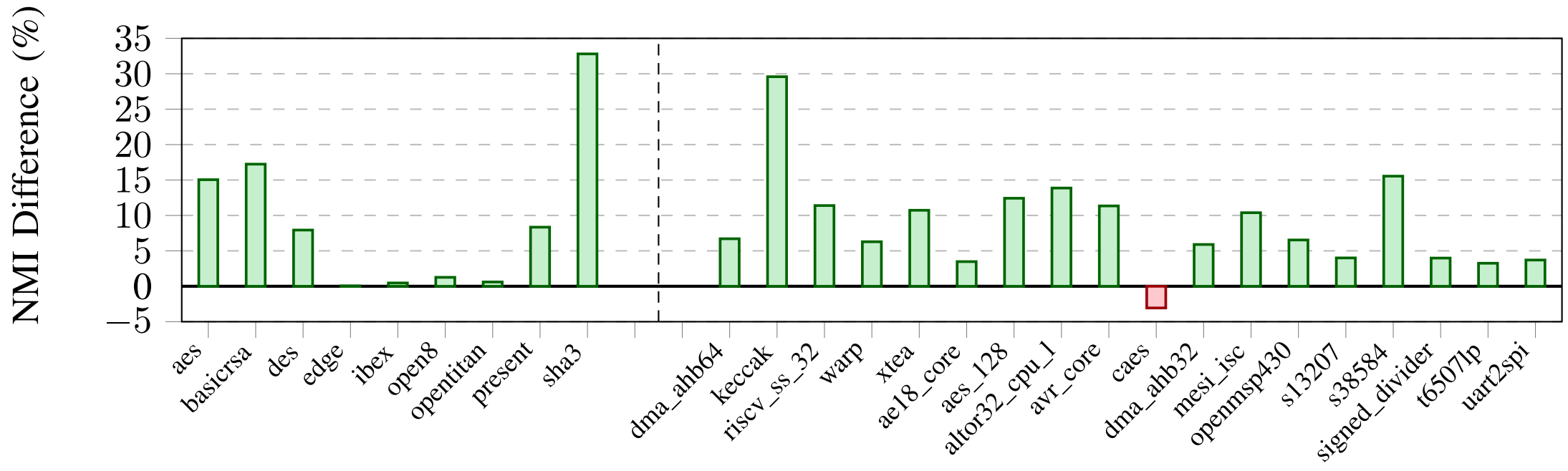

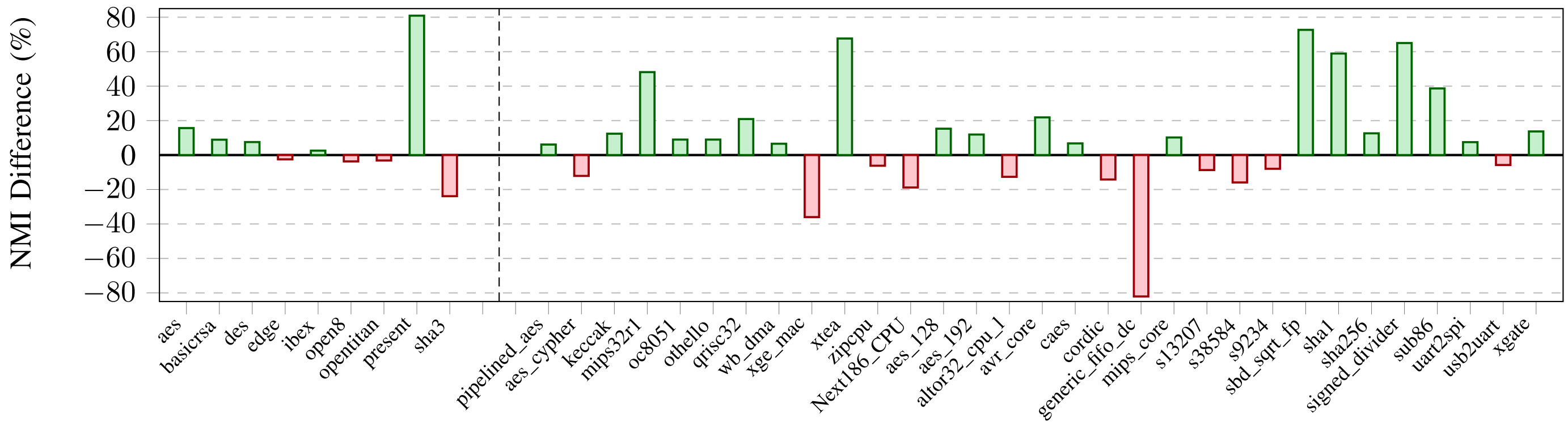
\begin{figure*}[ht]
\centering
\begin{tikzpicture}
\begin{axis}[
    ybar,
    width=0.99\textwidth,
    height=\bargraphheight,
    bar width=4.8pt,
    ymin=-85,
    ymax=85,
    ylabel={NMI Difference (\%)},
    ylabel style={xshift=4pt},
    xtick={0,...,41},
    xticklabels={aes,basicrsa,des,edge,ibex,open8,opentitan,present,sha3,,,pipelined\_aes,aes\_cypher,keccak,mips32r1,oc8051,othello,qrisc32,wb\_dma,xge\_mac,xtea,zipcpu,Next186\_CPU,aes\_128,aes\_192,altor32\_cpu\_l,avr\_core,caes,cordic,generic\_fifo\_dc,mips\_core,s13207,s38584,s9234,sbd\_sqrt\_fp,sha1,sha256,signed\_divider,sub86,uart2spi,usb2uart,xgate},
    x tick label style={rotate=45,anchor=east,font=\scriptsize},
    ymajorgrids=true,
    grid style={dashed},
    ytick={-80,-60,-40,-20,20,40,60,80},
    extra y ticks={0},
    extra y tick labels={0},
    extra y tick style={grid=major,grid style={solid,black,line width=0.8pt},tick style={draw=none}},
    xtick pos=bottom,
    ytick pos=left,
    axis line style={black},
    tick align=outside,
    enlarge x limits=0.02,
]

\addplot[ybar,bar shift=0pt,fill=positivefill,draw=positiveedge,line width=0.7pt] coordinates {
    (0,15.6585)
    (1,8.9195)
    (2,7.5426)
    (4,2.5712)
    (7,80.9049)

    (11,6.1024)
    (13,12.3961)
    (14,48.1370)
    (15,8.9821)
    (16,8.9838)
    (17,20.8933)
    (18,6.5764)
    (20,67.6432)

    (23,15.2951)  
    (24,11.9105)  
    (26,21.8716)  
    (27,6.7661)   
    (30,10.2053)  
    (34,72.6745)  
    (35,58.9244)  
    (36,12.6136)  
    (37,65.0159)  
    (38,38.6712)  
    (39,7.5003)   
    (41,13.7079)  
};

\addplot[ybar,bar shift=0pt,fill=negativefill,draw=negativeedge,line width=0.7pt] coordinates {
    (3,-2.4896)
    (5,-3.6910)
    (6,-3.1558)
    (8,-23.8804)

    (12,-12.1130)
    (19,-36.0130)
    (21,-6.2027)

    (22,-18.8273) 
    (25,-12.6387) 
    (28,-14.1839) 
    (29,-82.0707) 
    (31,-8.7164)  
    (32,-15.9099) 
    (33,-7.9738)  
    (40,-5.7705)  
};

\draw[black,densely dashed] (axis cs:9.5,-85) -- (axis cs:9.5,85);


\end{axis}
\end{tikzpicture}
\caption{NMI changes from \mytextsc{DANA} to \mytextsc{\projname}. The left portion of the plot shows the 9 original \mytextsc{DANA} designs; the right portion shows only benchmarks with $|\Delta \mathrm{NMI}| > 5\%$. Automatic clock gating is \textbf{disabled} for all these designs.}
\label{fig:nmi_wo_cg}
\end{figure*}


\mytextsc{DANA} achieves average NMI and purity values of 0.858 and 0.805, respectively, on the nine \mytextsc{DANA} benchmarks synthesized with ASAP7, compared with 0.88 and 0.86 reported in the original study~\cite{albartus2020dana}. We attribute the decrease in our replicated \mytextsc{DANA} results to the use of a different standard cell library, which affected logic synthesis and likely caused our synthesized netlists to differ slightly in gate composition and structure from the benchmarks in~\cite{albartus2020dana}. This small divergence is therefore expected.

Our proposed method, \mytextsc{\projname}, achieves average NMI and purity values of 0.951 and 0.900, respectively, on the \mytextsc{DANA} benchmarks, representing an approximate increase of 0.09 in NMI (a 10.8\% improvement) and 0.1 in purity over the \mytextsc{DANA} baseline. These substantial NMI and purity improvements demonstrate the effectiveness of \mytextsc{\projname}'s layout analysis for register identification and FF grouping. Despite the general score decrease associated with the ASAP7 library, \mytextsc{\projname}'s NMI values match or exceed the reported NMI values in~\cite{albartus2020dana}, and are higher than all \mytextsc{DANA} NMI values in our replication, for every design.

On the large benchmark set, \mytextsc{DANA} achieves average NMI and purity values of 0.871 and 0.778, respectively, and \mytextsc{\projname} achieves 0.896 and 0.778. The average NMI improvement is approximately 2.5\%, which is smaller than that observed on the \mytextsc{DANA} benchmark set. One reason is that \mytextsc{DANA} already achieves an NMI above 0.90 for 31 of the 60 designs in this dataset, leaving less room for improvement. However, among the remaining 29 designs, \mytextsc{\projname} increases NMI for 20. Of these 20 designs, 11 improve from an NMI below 0.90 with \mytextsc{DANA} to above 0.90 with \mytextsc{\projname}. 

In total, 37 designs exhibit an NMI increase, while 18 exhibit a decrease. The decreases are generally minuscule, with half having a magnitude less than 0.01 NMI. Fig.~\ref{fig:nmi_w_cg} further illustrates this trend: nearly all significant NMI changes are increases, and only one design shows a significant decrease. Interestingly, purity remains unchanged after applying \mytextsc{\projname}'s layout analysis, suggesting that \mytextsc{\projname} and \mytextsc{DANA} are similarly effective at avoiding heterogeneous register groupings, even though \mytextsc{\projname} yields more accurate register groupings overall.

For the benchmarks synthesized without clock gating, NMI undergoes a steep decline. Notably, on the \mytextsc{DANA} benchmark set, \mytextsc{DANA} achieves average NMI and purity values of 0.557 and 0.524, respectively, while \mytextsc{\projname} achieves 0.648 and 0.605. This corresponds to an improvement of approximately 0.09 in NMI and 0.08 in purity. However, the cryptocore PRESENT design accounts for most of the average NMI improvement. In contrast, for the clock gated benchmarks, the NMI improvements are more distributed across five designs.

Furthermore, on the large benchmark set, \mytextsc{DANA} achieves average NMI and purity values of 0.525 and 0.373, while \mytextsc{\projname} achieves 0.574 and 0.370, corresponding to a 9.3\% NMI increase over the \mytextsc{DANA} baseline. Fig. \ref{fig:nmi_wo_cg} reveals that this is by no means a uniform increase. For example, the NMI of the generic FIFO design declines by more than 0.80, and the NMI of xge\_mac decreases by almost 0.40. On the other hand, the NMI of xtea rises by 0.67, s9234 by 0.73, and six other designs by more than 0.20.  

Ultimately, the substantial NMI drop compared to the clock gated benchmarks indicates that \emph{clock gating is a major contributor to the strength of \mytextsc{DANA}'s register detection capabilities}. In the absence of clock gating, \mytextsc{\projname} still generally \textbf{outperforms} \mytextsc{DANA}, though its NMI varies considerably across designs. In contrast, the average purity difference between \mytextsc{DANA} and \mytextsc{\projname} is negligible (0.003), leading to the same conclusion that \mytextsc{\projname} improves overall register grouping accuracy without increasing grouping heterogeneity.

To analyze the statistical significance of the NMI improvements on the clock gated benchmarks, we perform paired t-tests on the NMI differences between \mytextsc{\projname} and \mytextsc{DANA}. The resulting \emph{p} values are 0.0339 and 0.00077 for the \mytextsc{DANA} and large benchmark sets, respectively. Both \emph{p} values are below 0.05, supporting the statistical significance of the observed NMI improvements for \mytextsc{\projname} compared to \mytextsc{DANA}. 

A comparison of the execution time is shown in Table~\ref{tab:runtime}. Although \mytextsc{\projname} has a slightly longer average runtime than \mytextsc{DANA}, it remains efficient, processing netlists with thousands of FFs and hundreds of thousands of gates within minutes.


\begin{table}[t]
    \centering
    \caption{Runtime comparison}
    \label{tab:runtime}
    \begin{tabular}{l c c c c c}
    \toprule
    \multirow{2}{*}{FF Count}
    & \multirow{2}{*}{\# Designs}
    & \multicolumn{2}{c}{\mytextsc{DANA}}
    & \multicolumn{2}{c}{\mytextsc{LaDA}} \\
    \cmidrule(lr){3-4}
    \cmidrule(lr){5-6}
    & & Avg. (s) & Max (s) & Avg. (s) & Max (s) \\
    \midrule
        $<500$                   & 17 & 0.42  & 1.70   & 0.46   & 1.54   \\
        $500$--$999$            & 12 & 1.43  & 6.47   & 1.47   & 6.80   \\
        $1{,}000$--$1{,}999$    & 15 & 3.32  & 9.13   & 3.49   & 8.73   \\
        $2{,}000$--$3{,}999$    & 13 & 12.43 & 48.25  & 12.35  & 47.88  \\
        $\geq 4{,}000$          & 12 & 87.69 & 397.57 & 118.01 & 475.09 \\
        \midrule
        \textbf{All designs}    & \textbf{69}
        & \textbf{18.67} & \textbf{397.57}
        & \textbf{23.98} & \textbf{475.09} \\
        \bottomrule
    \end{tabular}
\end{table}

\section{Discussion}


Our results demonstrate \mytextsc{\projname}'s effectiveness in enhancing register identification and improving clustering quality, particularly on designs where \mytextsc{DANA} has relatively lower grouping accuracy. The results also confirm \mytextsc{DANA} as a proficient baseline for register RE.

For the clock-gating experiments, the NMI decrease on benchmarks without clock gating reveals \mytextsc{DANA}'s dependency on this facet of logic synthesis. Although \mytextsc{\projname} generally improves upon \mytextsc{DANA} in this setting, the gains vary across designs. Nonetheless, as~\cite{albartus2020dana} notes, gated clock tree synthesis is a typical practice in digital design because of its benefits in reducing power consumption, so the clock-gated results remain practically relevant.

We also find that our theoretical model underlying \mytextsc{\projname}, in which FFs in the same register tend to be physically close and have similar predecessor-successor sets, holds true for most registers across most designs. Designs with NMI decreases tend to contain only one or two registers that do not follow this pattern, so the overall NMI reduction is usually small. By definition, these registers are typically small and of less interest for reverse engineering purposes. Furthermore, in many of these designs, \mytextsc{DANA} also has a high baseline NMI, leaving little room for further improvement. Future work could focus on these atypical cases to mitigate NMI losses, while preserving \mytextsc{\projname}'s NMI gains on other designs. 

Although \mytextsc{\projname} avoids explicit cell distance thresholds, it still includes several manually set parameters, including feature weights, thresholds, and parameters such as the \emph{k} value in the instances of k-nearest neighbors. As a whole, we believe these parameters are less dependent on physical scale than explicit, cell coordinate distance thresholds. We have also shown how the chosen parameters, on average, generalize well to a large set of varied benchmarks.


Furthermore, the final weights in the merge function indicate that predecessor similarity contributes more than successor similarity to register RE. This is consistent with our expectation that fan-in more directly captures the unique register-transfer conditions that determine the next state of a register, while fan-out is a less reliable predictor. 

\section{Conclusion}
Register RE provides insight into the dataflow and design of a netlist, and supports hardware security applications such as Trojan and IP piracy detection. Prior logical RE techniques primarily derive their heuristics from netlist information and largely overlook the physical design layout. To address this gap, we introduce \mytextsc{\projname}, a layout-aware register RE method that integrates layout cell coordinate information with FF dependency graph relationships. \mytextsc{\projname} achieves strong register detection performance, raising the average NMI from 0.86 to 0.95 on the \mytextsc{DANA} benchmarks, and improving register grouping across a broad set of open-source designs. In conclusion, \mytextsc{\projname} advances logic RE by incorporating physical design information into register detection.

\section*{Acknowledgments}
We thank the authors of \mytextsc{DANA} for open sourcing their technique. This research is funded by the NSF GRFP, under grant DGE2140739, and by the CMU ECE Ann and Martin McGuinn Graduate Fellowship.


\pagebreak 

\bibliographystyle{IEEEtran}
\bibliography{references}
\end{document}